\documentclass[12pt]{article}
\usepackage{amsmath}
\usepackage{latexsym}
\usepackage{amssymb}
\usepackage{fullpage}
\usepackage{tikz}
 \usepackage{colortbl}
\usepackage{epsfig}
\usepackage{graphics}
\usepackage{fancyhdr}

\newcommand{\qed}{\hfill \ensuremath{\Box}}
\def\dimo{\noindent\mbox{\sc proof: }}
\def\eproof{\rm\hspace*{\fill}$\Box$\vspace{10pt}}
\newtheorem{defin}{\bf Definition}[section]
\newtheorem{theo}[defin]{Theorem}
\newtheorem{lem}[defin]{Lemma}
\newtheorem{corol}[defin]{Corollary}
\newtheorem{prop}[defin]{Proposition}

\newtheorem{oss}[defin]{Remark}
\newtheorem{ex}[defin]{Example}
\def\cF{{\cal F}}
\def\cV{{\cal V}}
\def\cG{{\cal G}}
\def\cP{\mbox{\boldmath ${\cal P}$}}

\def\vr{{\varrho}}

\def\ccF{\mbox{\footnotesize \boldmath $\cF$}}

\def\cccF{\mbox{\boldmath $\cF$}}

\def\meglio {{\quad {\underset {\sim}  \succ}}}
\def\ccG{\mbox{\footnotesize \boldmath $\cG$}}
\def\cccG{\mbox{\boldmath $\cG$}}
\def\ccell{\mbox{\footnotesize \boldmath $\ell$}}
\def\cck{\mbox{\footnotesize \boldmath $k$}}
\def\cccell{\mbox{\boldmath $\ell$}}
\def\ccck{\mbox{\boldmath $k$}}
\def\cccm{\mbox{\boldmath $m$}}
\def\cccV{\mbox{\boldmath $\cV$}}
\def\ccj{\mbox{\footnotesize \boldmath $j$}}
\def\cccj{\mbox{\boldmath $j$}}
\def\cch{\mbox{\footnotesize \boldmath $h$}}
\def\ccch{\mbox{\boldmath $h$}}
\def\ccx{\mbox{\footnotesize \boldmath $x$}}
\def\cccx{\mbox{\boldmath $x$}}
\def\ccy{\mbox{\footnotesize \boldmath $y$}}
\def\cccy{\mbox{\boldmath $y$}}

\title{Anonymous,
non-manipulable, binary social choice
}
\author{Achille Basile\thanks{Corresponding author.}\\
Dipartimento di Scienze Economiche e Statistiche \\
 Universit\`a Federico II, 
80126 Napoli, Italy\\
E-mail: basile@unina.it\\
\\
Surekha Rao \\
School of Business and Economics\\
Indiana University Northwest,
Gary, IN 46408\\
E-mail: skrao@iun.edu\\
and\\
K. P. S. Bhaskara Rao\\
Department of Computer Information Systems\\
Indiana 
University Northwest, Gary, IN 46408\\
E-mail: bkoppart@iun.edu}

\begin{document}
\maketitle

\thispagestyle{empty}

\begin{abstract}
Let $V$ be a finite society whose members express weak orderings (hence also indifference, possibly) about two alternatives.  We show a simple representation formula that is valid for all, and only, anonymous, non-manipulable, binary social choice functions on $V$. The number of such  functions is $2^{n+1}$
if $V$  contains $n$ agents.
%{\color{red}If a  SCF for weak profiles is anonymous, it should be anonymous when it is restricted to strict profiles. So, a CSP SCF for weak profiles with range at least 3 cannot be anonymous.
%I think that we also know that range of SCF does not change when it is restricted to strict profiles.
%Yes. That is correct. We know that range of SCF does not change when it is restricted to strict profiles.

%also remind 

%{\color{red}\bigskip
%Before Bhaskara forgets: In case V is countable, every ASP csf corresponds to $(k_0, k_1, ...k_r)$ where each $ k_i$ is either finite or co-m for a finite m or $ \aleph_0 co-\aleph_0$ where is a finite inteteger. Also, there are only countably many ASP csf's.  I believe that this can be proved. We can think of this after a few days.  Similar statement can be made for any infinite cardinal also, I believe. Not sure if it is worth it. Leave this for the time being.}

%}
\end{abstract}




 \newpage
 
  \setcounter{page}{1}
 
\section{Introduction}
\lhead{\sc \scriptsize Representation of  anonymous...}
\rhead{\sc  \scriptsize Introduction  }

It is well known, after the celebrated Gibbard-Satterthwaite Theorem (\cite {G}, \cite{S}), that the two properties of anonymity and non-manipulability of a  social choice function, conflict every time the collective choice is from among a set  of at least three alternatives. Anonymity guarantees that all individuals of the collectivity are equally powerful in the social choice determination, whereas non-manipulability (or strategy-proofness) guarantees that telling the truth is strategically dominant for all individuals. The desirability of both properties resulted in a large literature
that considers, with three or more alternatives, social choice functions over restricted domains  or considers weaker properties to be satisfied by the collective choice  (a classical survey is \cite{B}).
At the same time, attention has been paid to the case, quite common in many important practical situations, in which the society has to decide by choosing between two alternatives $a$ and $b$. In the latter, binary, case,  if every voter  is asked to declare a strict preference, one has that the anonymous, non-manipulable social choice functions are all, and only, the  {\bf quota majority methods} (see \cite[Corollary of page 63]{M}), in the sense that  they are the $n+2$  functions $\mu_k$, with $k=0,1, 2, \dots, n+1$, defined as follows. For a profile $P=(P_v)_{v\in V}$ of preferences\footnote{In this case every voter $v$ is asked to declare either $P_v=a$ or $P_v=b$.}, the corresponding social choice  $\mu_k(P)$ is $a$ if the number of voters choosing $a$ is at least $k$, otherwise the collective choice is $b$.

%if the society $V$ has $n$ agents, there are $n+2$ anonymous, strategy-proof social choice functions. Moreover, they are {\bf quota majority methods}. 

\bigskip
This paper deals with the binary setting in which the voters are allowed to express  indifference also.  We shall describe in this setting all anonymous non-manipulable social choice functions by means of a representation formula that turns out to be  quite simple to describe and also is a direct extension of the quota majority rule (we call it {\bf extended quota majority}). As a straightforward corollary, we shall show that there are $2^{n+1}$  anonymous, non-manipulable, binary social choice functions  when the society has  $n$ voters. 

The fact that, due to the transition from strict to weak orderings, the number  of  anonymous, non-manipulable, binary social choice functions depends exponentially on the number of voters, rather than linearly, suggests  that obtaining  a sound representation theorem  for weak orderings is not an obvious task. To the best of our knowledge,   this problem was considered only recently in Lahiri and Pramanik 
 (\cite[Theorem 2]{LP}). In {\color{black} Section 4  we shall compare the two representations. We point out that ours is not only glaringly a plain extension of the quota majority method (indeed a sequence of majority rules), but also   can be made optimal in the sense of minimizing the set of the necessary parameters. The notion of extended quota majority method is quite intuitive. If one applies a quota majority method $\mu_{k_0}$ allowing for indifference, except for the cases in which the quota $k_0\in \{0, n+1\}$, where we get a constant collective choice, there will always be profiles for which the method does not provide a social choice. Then, the idea is to apply a further quota majority method $\mu_{k_1}$ but only to profiles for which $\mu_{k_0}$ has not given the value. Some more profiles will be covered, but also the application of the second quota may leave uncovered profiles. Then, one can apply a $\mu_{k_2}$, and so on. To ensure that this procedure stops, the sequence of quotas must involve at some point a quota $k_r\in\{0, n+1\}$. It turns out that this is the only way to produce anonymous, non-manipulable binary social choice functions. We go beyond this. We show that the choice of the quotas can be done in an optimal way, in the sense that the dimension of the vector $\ccck=(k_0, k_1,\dots,k_r)$  is minimum. We also characterize this by showing that it happens correspondingly to an up-down course of the sequence like, for example, the following
 $\dots <k_4<k_2<k_0<k_1<k_3<\dots$
 }

\bigskip
In \cite{KPS2} we have given a formula to represent all strategy-proof binary social choice functions, on an arbitrary set of voters (i.e. not necessarily finite, as we assume here), wherein voters are permitted to express indifference.
 We introduced a class of social choice functions, that we call $\psi${\it-type functions}, and shown 
that these are all, and only, the binary social choice functions that cannot be manipulated.  
We have seen that some (not all such functions) can have a simpler structure based on collections of committees (see \cite[Remark 2.7, Proposition 2.9]{KPS2})
potentially  within  the society. A significant example  is the simple majority rule (\cite[Proposition 4.3]{KPS2}). The results of this paper extend the exercise for the simple majority rule. We emphasize that they cannot be seen as straightforward consequences of the general representation theorem (\cite[Theorem 4.2]{KPS2}) obtained in \cite{KPS2}.
In particular this is true since our main result (Theorem \ref{main} below) is also  a uniqueness result of the representation of a non-manipulable
anonymous social choice function  by means of   some special extended quota majority methods that we call {\bf proper extended quota majority methods}.

\bigskip
The paper proceeds as follows. Next Section presents our results. The third Section contains all technical details of the needed proofs. Fourth Section concludes by comparing our representation formula with that proposed by Lahiri and Pramanik in \cite{LP}.

\section{Results}\label{risultati}
\lhead{\sc \scriptsize Representation of  anonymous...}
\rhead{\sc  \scriptsize Results  }

Let $V$ be a society of cardinality $n$. 
A profile $P = (P_v)_{v\in V}$ consists  of the declarations, denoted by $P_v$, of agent $v$'s preference between the alternatives  $a$ and $b$.  The collectivity $V$ will necessarily implement one of the two alternatives. 
Since we allow for indifference, the possibilities for agent $v$ are: to declare  preference for $a$, or  for $b$ or to declare indifference between $a$ and $b$.

A social choice function (scf, for short) $\phi$  is a
mapping $P \mapsto \phi(P)$, the value $\phi(P)$ being the alternative selected  as the social outcome corresponding to the profile $P$. Since throughout the paper we only deal with binary (i.e. only the two alternatives $a$ and $b$ are considered) scfs, we shall use scf to mean binary scf.

In order to ensure a fair consideration of the opinions of all agents, one may require anonymity of scfs. Non-manipulability may be required to prevent strategical false declarations. The formal, well established definitions, are:

\begin{defin}
A scf $\phi$ is:
\begin{itemize}
\item[] {\bf anonymous}, if \,\, $\phi(P)=\phi(P\circ \sigma)=\phi(\, (\,P_{\sigma(v)}\,)_{v\in V}\,),$ for every profile $P$ and for every permutation $\sigma$ of $V$.
\item[] {\bf non-manipulable}, if \,\, $ \phi(P_v, P_{-v}) \meglio_{P_v} \phi(Q_v, P_{-v})$, for every voter $v$, for every profile $P$, and for every weak ordering $Q_v$.
\end{itemize}
\end{defin}

We introduce now extended quota majority methods, denoted by $\phi_{\cck}$,  which can be described as follows. Let us call {\bf $r$-tuple}, for $1\le r\le n+1$, an ordered  tuple $\ccck=(k_0, k_1, \dots, k_r)$ of distinct elements from the set $\{0,1,\dots, n+1\}$ such that $k_{r}\in\{0, n+1\}$.
%be  an $(r+2)$-permutation  of the set $\{0,1,\dots, n, n+1\}$ \footnote{The terminology being not universally fixed, we specify that an $(r+2)$-permutation  of the set $\{0,1,\dots, n, n+1\}$ is defined as an injective mapping $\ccck: \{0, \dots, r+1\}\to \{0, \dots, n+1\}$.} such that $k_{r+1}\in\{0, n+1\}$. 
For a profile $P$, let $\lambda(P)$ be the smallest index $\lambda$ for which either at least $k_\lambda$ voters prefer $a$, 
%($|D(a,P)|\ge k_\lambda$)  
or at least $n+1-k_\lambda$ voters prefer $b$. 
%($|D(b,P)|\ge n+1-k_\lambda$)

Given these premises,  we give the following definition.
\begin{defin}\label{extended}
A scf is said to be an {\bf extended quota majority method} if  for some $r$-tuple $\ccck$  
%of the set $\{0,1,\dots, n, n+1\}$ such that $k_{r+1}\in\{0, n+1\}$
,  we have that the scf is defined as follows
\begin{equation*}
\phi_{\cck}(P) \overset{def}= \left\{ 
\begin{array} {ll}
a,   & \mbox{ if at least $k_{\lambda(P)}$ voters prefer } a,\\

 b,   & \mbox{ if at least $n+1-k_{\lambda(P)}$ voters prefer } b.\\
\end{array}
\right.
\end{equation*}
\end{defin}

\begin{oss} {\sl Notice that:
\begin{enumerate}
\item  When $P$ is  a strict profile, then obviously the index $\lambda(P)$ is zero, hence  $\phi_{\cck} (P)= \mu_{k_0}(P),$
for every $\ccck$, i.e. the extended method, restricted to strict profiles, gives back the original quota majority method.
\item It is immediate to recognize that, by Definition \ref{extended}, $k_0=n+1$ gives the collective choice being always $b$,
 irrespective of the profiles expressed by the collectivity. Analogously, if $k_0=0$ we get  always $a$ as the collective choice  irrespective of the profiles expressed by the collectivity. When $1\le k_0\le n$, the range of the collective choice is $\{a, b\}$.
\eproof \end{enumerate}
}\end{oss}

Now our first representation theorem can be promptly stated. 
\begin{theo}\label{first}
Extended quota majority methods $\phi_{\cck}$ are anonymous and non-manipulable. Moreover, every anonymous non manipulable binary social choice function is an extended quota majority  method for some $\ccck$.\end{theo}

 \begin{defin}
We say that the length of $\ccck$ is the smallest index $\lambda$ for which $k_\lambda\in\{0, n+1\}$.

%zero in case $\ccck$ is such that $k_0\in \{0, n+1\}$; whereas when $1\le k_0\le n$, the length of $\ccck$ is $r+1$ if    $r$ is the highest index for which $1\le k_r\le n$.
\end{defin}

The possibility of representing anonymous, strategy-proof scfs as extended quota majorities, does not ensure uniqueness of the representation. In order to achieve representations that are also unique, we introduce proper extended quota majorities.

\begin{defin}
%For an extended quota majority method to be {\bf proper}, means that it is onto\footnote{Hence, $1\le k_0\le n$.} and the sequence    $\ccck$ is proper, i.e. satisfies either one of the following   
 We say that an extended majority method is {\bf proper} if it satisfies one of the following
 {\bf up and down} conditions.
 
\bigskip
down-up:
$$
\begin{tabular}{ccc}
\hline
\multicolumn{1}{|c}{$0$} & \multicolumn{1}{|c}{$%
< k_{r-1}<...<k_{5}<k_{3}<k_{1}<k_{0}<k_{2}<k_{4}<k_{6}<...<$} & \multicolumn{1}{|c|}{%
$n+1=k_{r}$} \\ \hline
&  &  \\ \hline
\multicolumn{1}{|c}{$0=k_{r}$} & \multicolumn{1}{|c}{$%
<...<k_{5}<k_{3}<k_{1}<k_{0}<k_{2}<k_{4}<k_{6}<...<k_{r-1}<$} & \multicolumn{1}{|c|}{%
$n+1$} \\ \hline
\end{tabular}
$$

up-down:
$$
\begin{tabular}{ccc}
\hline
\multicolumn{1}{|c}{$n+1=k_{r}$} & \multicolumn{1}{|c}{$%
>...>k_{5}>k_{3}>k_{1}>k_{0}>k_{2}>k_{4}>k_{6}>...>k_{r-1}>$} & \multicolumn{1}{|c|}{%
$0$} \\ \hline
\multicolumn{1}{l}{} & \multicolumn{1}{l}{} & \multicolumn{1}{l}{} \\ \hline
\multicolumn{1}{|c}{$n+1$} & \multicolumn{1}{|c}{$%
>k_{r-1}>...>k_{5}>k_{3}>k_{1}>k_{0}>k_{2}>k_{4}>k_{6}>...>$} & \multicolumn{1}{|c|}{%
$0=k_{r}$} \\ \hline
\end{tabular}
$$
\end{defin}

\bigskip
The following figure, illustrates, in a society $V$ of 11 voters, relatively to the function $\phi_{\cck}$, the properness of the sequence $\ccck$: the solid line involves a down-up sequence $\ccck=(4,3,7,2,8,1,12)$ of length 6; the dashed line  an up-down sequence $\ccck=(4,5,3,7,1,12)$ of length 5.

$$
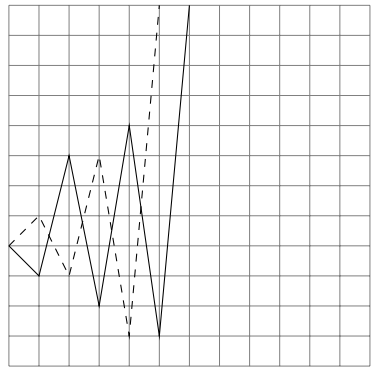
\begin{tikzpicture}[xscale=0.4, yscale=0.4]
\draw[help lines] (0,0) grid (12,12);
\draw (0,4) --(1,3) -- (2,7) -- (3,2) -- (4,8) -- (5,1) -- (6,12);  \draw [dashed](0,4) --(1,5) -- (2,3) -- (3,7) -- (4,1) -- (5,12) ;\end{tikzpicture}$$

\bigskip
The next one, illustrates, in a society $V$ of 11 voters, relatively to  $\phi_{\cck}$, the properness of the sequence $\ccck$: the solid line involves a down-up sequence $\ccck=(7,6,8,5,10,3,11,0)$ of length 7; the dashed line  an up-down sequence $\ccck=(4,7,3,9,0)$ of length 4.

$$
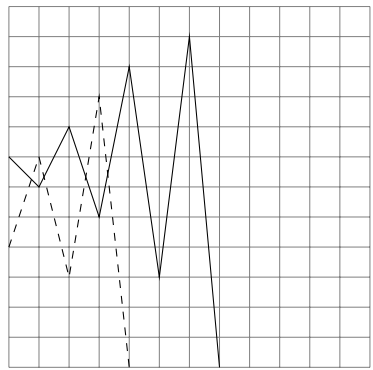
\begin{tikzpicture}[xscale=0.4, yscale=0.4]
\draw[help lines] (0,0) grid (12,12);
\draw (0,7) --(1,6) -- (2,8) -- (3,5) -- (4,10) -- (5,3) -- (6,11) -- (7,0) ;  
\draw [dashed](0,4) --(1,7) -- (2,3) -- (3,9) -- (4,0);  \end{tikzpicture}$$

Our main results is the following.
\begin{theo}\label{main}
 For every onto binary social choice function $\phi$ which is  anonymous, and non-manipulable there exists one and only one  proper extended quota majority method $\phi_{\cck}$ such that $\phi=\phi_{\cck}$.
\end{theo}

Because of Theorems \ref{first} and  \ref{main}, the following is now obvious.
\begin{corol}
The onto scfs that are  anonymous, and strategy-proof are all, and only, the proper extended quota majority methods.
\end{corol}

To determine the cardinality of the class of all anonymous, non-manipulable, binary social choice functions in a society with $n$ agents, we can count the extended quota majority methods. Indeed we count that there are $2^n$  anonymous, non-manipulable scfs corresponding to the collective choice $b$ for the unanimous indifference. Symmetrically, there are $2^n$ anonymous, non-manipulable scfs corresponding to the collective choice $a$ for the unanimous indifference.

To see the first statement, let us take a subset $J\subseteq \{1, \dots, n\}$. If $J$ is empty, let us associate $J$ with the constant scf $b$. Suppose $J$ is nonempty and has cardinality $r$. In this case we associate $J$ with the scf  $\phi_{(k_0, k_1, \dots, k_{r-1}, n+1)}$ where the proper $\ccck$ presents:

$$k_{r-1}= \min J,\, k_{r-2}=\max J\setminus \{k_{r-1}\}, \, k_{r-3}=\min J\setminus \{k_{r-1}, k_{r-2}\}, $$ $$ k_{r-4}=\max J\setminus \{k_{r-3}, k_{r-2}, k_{r-1}\}, \,\dots, 
k_0= \left\{ 
\begin{array} {ll}
\max J\setminus \{k_1, \dots, k_{r-1}\},   & \mbox{ if  } k_1 \mbox { is a min} \\

\min J\setminus \{k_1, \dots, k_{r-1}\},   & \mbox{ if  } k_1 \mbox { is a max}.\\
\end{array}
\right.$$ Theorems \ref{first} and  \ref{main} guarantees that the correspondence defined above is a bijection.
Hence by symmetry, we have proved the following.

\begin{corol}\label{counting}
There are $2^{n+1}$ anonymous, non-manipulable, binary social choice functions if $n$ voters choose between two alternatives, being allowed to express indifference.
\end{corol}

\begin{oss}\label{nest} {\sl A few comments are in order.

\begin{enumerate}
 \item Definition \ref{extended} con be formally given  with reference to  an arbitrary sequence $\cccj=(j_0, j_1, ...)$. The only needed condition is that at least one of its values belongs to $\{0, n+1\}$.  When $1\le j_0\le n$, if $r$ is the smallest index for which  $j_{r}\in\{0,n+1\}$, the original sequence $\cccj$ and the truncated sequence   $(j_0, j_1, ..., j_r)$ give rise to the same scf. 
 In this  case, $j_{r}=n+1$ corresponds to assign $b$ to the profile where  all agents are unanimously indifferent. An analogous comment applies when $j_{r}=0$ replaces $n+1$, in that case $a$ replaces $b$.

\item  Let $\phi_{\cch}, \,\, {\ccch}=(h_\lambda)_{\lambda\in \{0,\dots, \}}$ be defined with the help of an arbitrary sequence. If we have $h_\beta\ge h_\gamma\ge h_\alpha$ for indices $\alpha, \beta <\gamma$, then $\phi_{\cch}$ does not change if we remove $h_\gamma$ from the sequence $\ccch$ \footnote{ It is sufficient to observe that the index of a profile $P$ does not change if we remove $h_\gamma$. Indeed it is obvious that the index $\lambda(P)$ cannot be $\gamma$.
 }. The deletion of $h_\gamma$ is possible even if we have $h_\beta\le h_\gamma\le h_\alpha$  (or $h_\beta=h_\gamma$)  for indices $\alpha, \beta <\gamma$.
 \item By applying to the truncated sequence $(j_0, j_1, ..., j_r)$ above the deletion of the repeated indices, we recognize that the scf given by the formula of Definition \ref{extended} applied to an arbitrary $\cccj$ produces an extended quota majority method.
\eproof \end{enumerate}
}\end{oss}

 \begin{defin}
We say that a representation of a scf $\phi$ as an extended quota majority method is minimal if it has minimum length among all such representations of $\phi$.
\end{defin}
We close this Section by observing that
\begin{corol}\label{minimal}
A representation is proper if and only if it is minimal.
\end{corol}

The following example illustrates some of the concepts and  it is relevant for Remark \ref{sempre per bhaskara}
 \begin{ex}\label{per bhaskara}
{\sl In this example, to modify the status quo (say, $b$) to the new status (say, $a$), the society $V=\{1, \dots,11\}$ needs that at least two individuals wish doing that. However, if such individuals are no more than four, it is needed also that the voters in favor of maintaining the status quo are less than seven.

The model for this situation is  the scf $\phi$ defined as follows:
%, with reference to a society \{1, \dots, 11\}:

\bigskip
$\phi(P) = a$ if either $|D(a,P)|\ge 5$ or $2\le |D(a,P)|< 5  \,\&\, |D(b,P)|<7$. In all the other cases $\phi(P)=b$.

\bigskip

According to Remark \ref{nest}, for such a $\phi$ possible defining sequences are $(5, 2, 7, 12)$, \, $(5, 2, 9, 12)$,\, $(5, 2, 12)$, the latter being the proper representation.\eproof}
\end{ex}

\section{Proofs}
\lhead{\sc \scriptsize Representation of  anonymous...}
\rhead{\sc  \scriptsize Proofs  }

The investigation of the scfs which are non-manipulable relies (see \cite{LS}, \cite{LP}, \cite{KPS2}) on the so-called {\bf committees} and their {\bf duals}. A committee is, by definition,  a nonempty, closed under superset, familiy of coalitions that can be formed in the society $V$. 
We are particularly interested in families $\cG_k=\{E\subseteq V: |E| \geq k\}$
 for  $k=1, \dots, n$. The superset closed family dual\footnote{We remind from \cite{KPS2}, where the notion has been introduced, that the dual of a committee $\cF$ is the committee $\cF^\circ\overset{def}=\{E\subseteq V: V\setminus E\notin\cF\}.$} to $\cG_k$ is  $\cG_k^\circ=\{E\subseteq V: |E| \geq n+1-k\}$, i.e. $\cG_{n+1-k}$.

Let  $\cccV=\{\cG_k: k=1,\dots, n\}$ be the set of all such committees on $V$ that we refer to as {\bf committees of
of cardinal type $k$}. A synonym is {\bf superset closed family} (SSCF, for short) {\bf of cardinal type $k$}. It will be convenient to consider  the power set of $V$ and the empty subset of the power set of $V$ also as committees. They can be considered as of  cardinal type   respectively zero ($\cG_0=2^V$) and $n+1$ ($\cG_{n+1}=\O$). They are also dual to each other.

\smallskip
We shall also consider, given a subsets $I$ of $V$ with cardinality $\ell, \, (0\leq\ell<n)$, SSCFs  $\cF$ on $I^c=V\setminus I$ of cardinal type. If we suppose that  the type of $\cF$ is $k \,$ (necessarily we have $1\leq k\leq n-\ell$), the corresponding dual (with respect to $I^c$) family is also of cardinal type and has type  $m= n-\ell-k+1.$

\bigskip
By comparing Definition \ref{extended} with \cite[Remark 3.3]{KPS2}, it is evident that:
\begin{oss}
{\sl An onto extended quota majority method $\phi_{\cck}$,  $\ccck=(k_0, k_1, ..., k_r)$, is a scf of the form (see \cite[Remark 3.3]{KPS2}) $\phi_{\ccF, \, x}$ , where $x$ is either $a$ or $b$ according to the fact that the first $k_i\notin\{1, \dots, n\}$ is either $0$ or $n+1$, and the collection $\cccF$  consists of the committees $\cG_{k_0}, \dots, \cG_{k_{i-1}}$ .

%untill $k_0, k_1, \dots $ belong to $\{1, \dots, n\}$.
}\end{oss}
The above Remark and \cite[Proposition 3.4]{KPS2}  tell us that extended quota majorities are non-manipulable scfs. Anonymity being obvious, we have the first part of Theorem \ref{first}.   
We shall prove the second part, namely that every anonymous, non manipulable scf is an extended quota majority method, in subsection \ref {prova1}. The fact that extended quota majority methods admit proper representations and that the proper representation is unique for onto scfs, is discussed in the subsection that follows.

%\newpage
\subsection{Existence and uniqueness of proper representations}
\lhead{\sc \scriptsize Representation of  anonymous...}
\rhead{\sc  \scriptsize Existence and uniqueness of proper representations  }
Throughout the sequel of the paper we adopt the following notation: by $D(a, P), D(b, P),$ and $I(P)$ we denote the subsets of $V$ consisting of voters that, respectively, choose $a, b$ or are indifferent between the two alternatives.

 \begin{prop}\label{33}
 Let $\phi_{\ccj}, \,\, {\cccj}=(j_\lambda)_{\lambda\in \{0,\dots, r\}}$, be an onto extended quota majority method. There is one and only one proper (sub)sequence $\ccck$ of $\cccj$ such that $\phi_{\ccj}=\phi_{\cck}$.
 \end{prop}
 
 \dimo
 
 We shall first show the existence of $\ccck$. Without loss of generality we can assume that 
 %$$\phi=\phi_{k_0, j_1, \dots, j_r, j_{r+1}},$$  where 

$  \{j_0, j_1, \dots, j_{r-1}\}\subseteq\{1, \dots, n\}$.

We  shall describe a procedure that, through the deletion of suitable indices $j$'s, produces the proper representation $\phi_{\cck}$ of $\phi_{\ccj}$. The initial element $k_0$ of $\ccck$ is set to be $j_0$.

\medskip
We discuss the case  $j_1>k_0$. The case $j_1<k_0$ can be discussed in a similar way.

\medskip
Let us partition the sequence defining $\phi_{\ccj}$ as illustrated in the following figure

$$\begin{tabular}{cccccc}
& \multicolumn{1}{|c}{$j_{1},...,j_{i_{1}-1}$} & \multicolumn{1}{|c}{} & 
\multicolumn{1}{|c}{$j_{i_{2}},...,j_{i_{3}-1}$} & \multicolumn{1}{|c}{} & 
\\ \hline
$k_{0}$ &  &  &  &  & $...$ \\ \hline
&  & \multicolumn{1}{|c}{$j_{i_{1}},...,j_{i_{2}-1}$} & \multicolumn{1}{|c}{}
& \multicolumn{1}{|c}{$j_{i_{3}},...,j_{i_{4}-1}$} & \multicolumn{1}{|c}{}%
\end{tabular}$$
The indices $i_1, i_2, \dots $ are  defined as follows:

$i_1=\min \{i>1: j_i< j_0\}$; \,\, $i_2=\min \{i>i_1: j_i>j_0\}$;\,\,
$i_3=\min \{i>i_2: j_i< j_0\}$; \,\,
$\dots$

One can adopt the usual convention that the minimum of the empty set is $\infty$ and once $i_h=\infty$ the sequence of indices $i$'s stops. The top row contains values of (cardinalities) $j$'s bigger than $k_0$, the bottom one smaller than $k_0$

Let $k_1$ be defined as the maximum of $\{j_{1},...,j_{i_{1}-1} \}$. It is a straightforward calculation to verify that $\phi_{\ccj}=\phi_{k_0, k_1,j_{i_{1}},...,j_{i_{2}-1}, \dots, j_r}$.

Let $k_2$ be defined as the minimum of $\{j_{i_{1}},...,j_{i_{2}-1} \}$. It is a straightforward calculation to verify that $\phi_{\ccj}=\phi_{k_0, k_1, k_2, j_{i_{2}},...,j_{i_{3}-1}, \dots, j_r }$.

If one of the values in the set $\{ j_{i_{2}},...,j_{i_{3}-1}\}$ is smaller than $k_1$, due to Remark \ref{nest}, such value can be deleted without modifying the scf $\phi_{\ccj}$. Therefore, we can assume we are in a situation like this:
$$k_2<k_0<k_1<\{ j_{i_{2}},...,j_{i_{3}-1}\}.$$
Let $k_3$ be defined as the maximum of $\{j_{i_{2}},...,j_{i_{3}-1} \}$. It is a straightforward calculation to verify that $\phi_{\ccj}=\phi_{k_0, k_1,k_2, k_3, j_{i_{3}},...,j_{i_{4}-1}, \dots, j_r}$.

If one of the values in the set $\{ j_{i_{3}},...,j_{i_{4}-1}\}$ is bigger than $k_2$, due to Remark \ref{nest}, such value can be deleted without modifying the scf $\phi_{\ccj}$. Therefore, we can assume we are in a situation like this:
$$\{ j_{i_{3}},...,j_{i_{4}-1}\}<k_2<k_0<k_1<k_3.$$
Let $k_4$ be defined as the minimum of $\{ j_{i_{3}},...,j_{i_{4}-1} \}$. It is a straightforward calculation to verify that $\phi_{\ccj}=\phi_{k_0, k_1,k_2, k_3, k_4, j_{i_{4}},...,, \dots, j_r}$.
We continue this way till we produce a proper $\ccck$ giving the same scf.
%
%We proceed similarly. The partition looks like
%$$
%\begin{tabular}{cccccc}
%& \multicolumn{1}{|c}{} & \multicolumn{1}{|c}{$j_{i_{1}},...,j_{i_{2}-1}$} & 
%\multicolumn{1}{|c}{} & \multicolumn{1}{|c}{$j_{i_{3}},...,j_{i_{4}-1}$} & 
%\multicolumn{1}{|c}{} \\ \hline
%$k_{0}$ &  &  &  &  & $...$ \\ \hline
%& \multicolumn{1}{|c}{$j_{1},...,j_{i_{1}-1}$} & \multicolumn{1}{|c}{} & 
%\multicolumn{1}{|c}{$j_{i_{2}},...,j_{i_{3}-1}$} & \multicolumn{1}{|c}{} & 
%\multicolumn{1}{|c}{}%
%\end{tabular}
%$$
%The indices $i_1, i_2, \dots $ are:
%
%$i_1=\min \{i>1: j_i> j_0\}$

%$i_2=\min \{i>i_1: j_i<j_0\}$

%$i_3=\min \{i>i_2: j_i> j_0\}$

%$\dots$

%The cardinalities of the canonical representation: 

% $k_1=\min\{j_{1},...,j_{i_{1}-1} \}$. 

% $k_2=\max\{j_{i_{1}},...,j_{i_{2}-1} \}$.

% We can assume we are in a situation like this:
%$$\{ j_{i_{2}},...,j_{i_{3}-1}\}<k_1<k_0<k_2.$$
 %$k_3=\min\{j_{i_{2}},...,j_{i_{3}-1} \}$. 

%We can assume we are in a situation like this:
%$$k_3<k_1<k_0<k_2<\{ j_{i_{3}},...,j_{i_{4}-1}\}.$$
% $k_4=\max \{ j_{i_{3}},...,j_{i_{4}-1} \}$. 

%Etc.
 
 \bigskip
 We shall now see the uniqueness.
For a scf $\phi$ giving value $b$ for unanimous indifference,  suppose that we have two proper representations $\phi_{\cck}=\phi_{k_0, k_1, \dots, k_{r-1}, n+1}$
and $\phi_{\cch}=\phi_{h_0, h_1, \dots, h_{s-1}, n+1}$. 
We shall show first that $k_0=h_0$, then $k_1=h_1$, $k_2=h_2$, and so on. Afterwords we show that also the lengths coincide. We shall write $n-k+1$ as $k^\circ$.

\bigskip
To see $k_0=h_0$:

Suppose, without loss of generality, that $k_0<h_0$. Let $P$ be a profile with $|D(a, P)|=k_0$ and $|D(b, P)|=h_0^\circ$. By using the $\ccck$-representation, we get that the social choice is $a$. By using the $\ccch$-representation, we get that the social choice is $b$ and this is a contradiction.

\bigskip
To see $k_1=h_1$:

Suppose, without loss of generality, that $k_1<h_1$. We shall consider the following three cases:
$k_0<k_1<h_1$, \, $k_1<k_0<h_1$, \, $k_1<h_1<k_0$, obtaining for everyone a contradiction to the assumption that $\phi_{\cck}=\phi_{\cch}.$

\bigskip
When $k_0<k_1<h_1$, since $\ccck$ is proper, we have necessarily $\quad \dots<k_2< k_0<k_1<h_1$. 
Let $P$ be a profile with $|D(a, P)|=k_0-1$ and $|D(b, P)|=h_1^\circ $. 
Along $\ccck$ the index of $P$ is two and the value of $\phi(P)$ is $a$.
Along $\ccch$ the index of $P$ is one and the value of $\phi(P)$ is $b$, a contradiction.

\bigskip
For the other two cases $k_1<k_0<h_1$, \,and \, $k_1<h_1<k_0$, 
let us consider  a profile $P$ with $|D(a, P)|=k_1$ and $|D(b, P)|=k_0^\circ -1$. By using the $\ccck$-representation, we get that the social choice is $a$. By using the representation $\phi=\phi_{k_0, h_1, \dots, h_{s-1}, n+1}$, we get that the social choice is $b$ and this is a contradiction.

To show that $\phi_{k_0, h_1, \dots, h_{s-1}, n+1} (P)=b$, in the  case $k_1<k_0<h_1$, since $k_0^\circ-1\ge h_1^\circ$, clearly $\phi_{k_0, h_1, \dots, h_{s-1}, n+1} (P)=b$.
In the case $k_1<h_1<k_0$,  since the index of the profile $P$ cannot be zero, first  notice that it is also not one. Indeed, not only $|D(a, P)|<h_1$, but also $|D(b, P)|=n-k_0 <n- h_1 <h_1^\circ$. 
Since $\ccch$ is proper, we have $k_1<h_1<k_0<h_2$, and  the index of $P$ is two: 
$|D(b, P)|=n-k_0>n-h_2\Rightarrow |D(b, P)|\ge h_2^\circ$. 

\bigskip
We can now suppose, without loss of generality, that we are in a situation like the following

$$\dots <k_5<k_3<k_1<k_0<k_2<k_4<k_6<\dots $$

$$\dots <h_5<h_3<k_1<k_0<h_2<h_4<h_6<\dots $$

\bigskip
After having proved, for $i\leq \min\{r-1,s-1\}$, that $k_\lambda=h_\lambda$ for $\lambda<i$, we prove by contradiction that $k_i=h_i$. As in the previous steps, let us assume that $k_i<h_i$.

We have 

$$\phi_{k_0, k_1, \dots,k_{i-1}, k_i, \dots, k_{r-1}, n+1}=\phi_{k_0, k_1, \dots,k_{i-1}, h_i, \dots, h_{s-1}, n+1}$$
and one of the following two cases.

Case 1:
$$
\begin{tabular}{lll}
\hline
\multicolumn{1}{|l}{$k_{i+1}$} & \multicolumn{1}{|l}{$%
<k_{i-1}<...<k_{1}<k_{0}<k_{2}<...<k_{i-2}<$} & \multicolumn{1}{|l|}{$k_{i}$}
\\ \hline
&  &  $\wedge $ \\ \hline
\multicolumn{1}{|l}{$h_{i+1}$} & \multicolumn{1}{|l}{$%
<k_{i-1}<...<k_{1}<k_{0}<k_{2}<...<k_{i-2}<$} & \multicolumn{1}{|l|}{$h_{i}$}
\\ \hline
\end{tabular}
$$

\bigskip
 Let $P$ be a profile with $|D(a, P)|=k_{i-1}-1$ and $|D(b, P)|=h_i^\circ$. By using the representation $\phi=\phi_{k_0, k_1, \dots,k_{i-1}, h_i, \dots, h_{s-1}, n+1}$, the index of the profile is $i$ and the social choice is $b$. 

By using the representation $\phi=\phi_{k_0, k_1, \dots, k_{r-1}, n+1}$, we get that the social choice is $a$ and this is a contradiction. To show that $\phi_{k_0, k_1, \dots, k_{r-1}, n+1} (P)=a$, notice that $i<r-1$ otherwise 
we would have both $k_{r-1}$ and $k_{r}=n+1$ on the right of $k_0$ against the properness of the representation. We see now that with respect to the representation $\phi=\phi_{k_0, k_1, \dots, k_r, n+1}$ the profile has index $i+1$ and the social choice is $a$.

\bigskip
Case 2:
$$\begin{tabular}{lll}
\hline
\multicolumn{1}{|l}{$k_{i}$} & \multicolumn{1}{|l}{$%
<k_{i-2}<...<k_{1}<k_{0}<k_{2}<...<k_{i-1}<$} & \multicolumn{1}{|l|}{$k_{i+1}
$} \\ \hline
$\wedge $ &  &  \\ \hline
\multicolumn{1}{|l}{$h_{i}$} & \multicolumn{1}{|l}{$%
<k_{i-2}<...<k_{1}<k_{0}<k_{2}<...<k_{i-1}<$} & \multicolumn{1}{|l|}{$h_{i+1}
$} \\ \hline
\end{tabular}
$$

\bigskip
 Let $P$ be a profile with $|D(a, P)|=k_{i}$ and $|D(b, P)|=k_{i-1}^\circ-1$. By using the representation $\phi=\phi_{k_0, k_1, \dots, k_{r-1}, n+1}$, we get that the social choice is $a$ (index is $i$). By using the representation $\phi=\phi_{k_0, k_1, \dots,k_{i-1}, h_i, \dots, h_{s-1}, n+1}$, we get that the social choice is $b$ and this is a contradiction. To show that $\phi_{k_0, k_1, \dots,k_{i-1}, h_i, \dots, h_{s-1}, n+1} (P)=b$, since the index of the profile $P$ with respect to this representation cannot be less than $i$, first  notice that it is also not $i$. Indeed, not only $|D(a, P)|<h_i$, but also $|D(b, P)|=n-k_{i-1} <n- h_i <h_i^\circ$. 
 
 If $s-1> i$, along the sequence $k_0, h_1, \dots, h_{s-1}$, the index of $P$ is $i+1$: 
$|D(b, P)|=n-k_{i-1}>n-h_{i+1}\Rightarrow |D(b, P)|\ge h_{i+1}^\circ$.  If $s-1=i$, then the index is $s$. In both cases the value of the social choice is $b$ as announced.

\bigskip
The indices $r$ and $s$ coincide.

Suppose $r>s$, then necessarily we have a situation as the following

$$
\begin{tabular}{lll}
\hline
\multicolumn{1}{|l}{$k_{r-1}<...<k_{s+1}<$} & \multicolumn{1}{|l}{$%
k_{s-1}<k_{s-3}<...<k_{1}<k_{0}<k_{2}<...<k_{s-2}<$} & \multicolumn{1}{|l|}{$%
k_{s}<...< k_{r}=n+1$} \\ \hline
&  &  \\ \hline
\multicolumn{1}{|l}{} & \multicolumn{1}{|l}{$%
k_{s-1}<k_{s-3}<...<k_{1}<k_{0}<k_{2}<...<k_{s-2}<$} & \multicolumn{1}{|l|}{$%
h_{s}=n+1$} \\ \hline
\end{tabular}
$$
If we take a profile $P$ $|D(a, P)|=k_{s-1}-1$ and $|D(b, P)|=0$. By using the representation $\phi=\phi_{k_0, k_1, \dots, k_{r-1}, n+1}$, we get that the social choice is $a$ (index $s+1$). By using the representation $\phi=\phi_{k_0, k_1, \dots,k_{s-1}, n+1}$, we get that the social choice is $b$ and this is a contradiction. \qed

\bigskip

{\sc proof} {\bf of Corollary \ref{minimal}}:

Let us take a representation $\phi_{k_0, k_1, \dots, k_{r-1}, n+1}$ of $\phi$ whose length is minimum and suppose that it is not proper. Let us suppose that $k_1>k_0$, since a similar argument can be given if the reverse inequality holds true. Since the representation is not proper, we shall have for some index $i\ge 1$ either the  condition:

$$k_{i-1}< ...<k_{2}<k_{0}<k_{1}<...<k_{i} \mbox{ and } k_{i+1}>k_{i-1},$$
or the condition
$$k_{i}< ...<k_{2}<k_{0}<k_{1}<...<k_{i-1} \mbox{ and } k_{i+1}<k_{i-1}.$$

In the first case the deletion of the smallest between $k_i$ and $k_{i+1}$ still produces a representation of $\phi$. In the second case the deletion of the largest between $k_i$ and $k_{i+1}$ still produces a representation of $\phi$. Hence $\phi_{k_0, k_1, \dots, k_{r-1}, n+1}$ is not of minimum length.
For the converse, namely, to show that every proper representation is minimal,  if $\phi_{k_0, k_1, \dots, k_{r-1}, n+1}$ is a proper representation of $\phi$, by the uniqueness  Theorem, it must be necessarily minimal. \footnote{Minimal representation obviously exist and we have just seen they are proper.}\qed

 \subsection{Subsets of profiles}
\lhead{\sc \scriptsize Representation of  anonymous...}
\rhead{\sc  \scriptsize Subsets of profiles}
We shall introduce here some useful subsets of the set  $\cP$ of all profiles. 

\bigskip

\begin{defin}\label{13d}
Let the natural numbers $\ell, k, m$ be such that $\,\, 0\leq\ell<n, \,\, 1\leq k\leq n-\ell, \\ m=n-\ell-k+1.$\footnote{The fact that the numbers $\ell, k$ and $m$ (possibly indexed) will be always such that $0\leq\ell<n, \qquad 1\leq k\leq n-\ell,\qquad m=n-\ell-k+1,$ will be assumed throughout the sequel of the paper.
} Let $I$ be a subset of $V$ with cardinality $\ell$, and  $\cF$  a SSCF on $I^c$
of cardinal type $k$ (with respect to $I^c$).

\begin{itemize}
\item[1.] The sets $\cP_a (\ell, k)$ and $\cP_b (\ell, k)$:

$$P\in \cP_a (\ell, k) \overset{def}\Leftrightarrow 
 \left\{ 
\begin{array} {ll}
|D(a,P)|\geq k ,   & \\

|D(b,P)|< m.   & \\
\end{array}
\right.
$$

$$
P\in \cP_b (\ell, k)\overset{def}\Leftrightarrow 
 \left\{ 
\begin{array} {ll}
|D(a,P)|< k,   & \\

|D(b,P)|\geq m.   & \\
\end{array}
\right.
$$

\item[2.] The sets $\cP_a(I, \cF)$ and $\cP_b(I, \cF)$:

$$P\in \cP_a(I, \cF) \overset{def}\Leftrightarrow 
 \left\{ 
\begin{array} {ll}
|D(a,P)\cap I^c|\geq k ,   & \\

D(b,P)\subseteq I^c.   & \\
\end{array}
\right.
$$

$$
P\in \cP_b(I, \cF)\overset{def}\Leftrightarrow 
 \left\{ 
\begin{array} {ll}
D(a,P)\subseteq I^c,   & \\

|D(b,P)\cap I^c|\geq m.   & \\
\end{array}
\right.
$$
In the particular case of $I=\O$ (i.e. $\ell=0$) we shorten $\cP_a(\O, \cG_k)$ as $\cP_a(\cG_k)$ and $\cP_b(\O, \cG_k)$ as $\cP_b(\cG_k)$.\footnote{ Hence:\quad 
$P\in\cP_a(\cG_k)$ means $|D(a,P)|\geq k$, and 
$P\in\cP_b(\cG_k)$ means $|D(b,P)|\geq n-k+1$.}
\item[3.] $\cP(\ell, k)\overset{def}=\cP_a (\ell, k)\cup \cP_b (\ell, k)$, \,\, and \,\, $\cP(\cG_k)\overset{def}=\cP_a (\cG_k)\cup \cP_b (\cG_k)$.
\end{itemize}
\end{defin}

\bigskip
Trivially, the sets $\cP_a$ and $\cP_b$ are disjoint. 

The profiles belonging to $\cP_a(\ell, k)$ have a stucture that can be also described as in the next proposition. Indeed, if we take a profile $P\in\cP_a(\ell, k)$, we can certainly fix a subset of $D(a, P)$ with cardinality $k$. Let us call $V_1$ such a set. Call $V_2$ a subset of $V\setminus V_1$ which is a superset of $D(b,P)$ with cardinality $n-\ell-k=m-1$. Finally, define $I=V\setminus (V_1\cup V_2)$ (the cardinality of $I$ is $\ell$). Hence, the profile $P$ looks like in the following figure 
$$
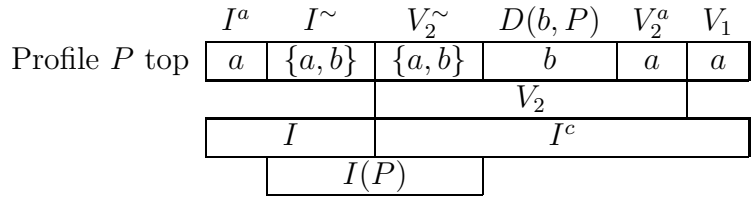
\begin{tabular}{lccccccl}
& $I^{a}$ & $I^{\sim }$ & $V_{2}^{\sim }$ & $D(b,P)$ & $V_{2}^{a}$ & $V_{1}$
&  \\ \cline{2-7}
Profile $P$ top & \multicolumn{1}{|c}{$a$} & \multicolumn{1}{|c}{$\{a,b\}$}
& \multicolumn{1}{|c}{$\{a,b\}$} & \multicolumn{1}{|c}{$b$} & 
\multicolumn{1}{|c}{$a$} & \multicolumn{1}{|c}{$a$} & \multicolumn{1}{|l}{}
\\ \cline{2-7}\cline{4-6}
&  &  & \multicolumn{3}{|c}{$V_{2}$} & \multicolumn{1}{|c}{} &  \\ 
\cline{2-7}
& \multicolumn{2}{|c}{$I$} & \multicolumn{4}{|c}{$I^{c}$} & 
\multicolumn{1}{|l}{} \\ \cline{2-7}\cline{3-4}
&  & \multicolumn{2}{|c}{$I(P)$} & \multicolumn{1}{|c}{} &  &  &  \\ 
\cline{3-4}
& \multicolumn{1}{l}{} & \multicolumn{1}{l}{} & \multicolumn{1}{l}{} & 
\multicolumn{1}{l}{} & \multicolumn{1}{l}{} & \multicolumn{1}{l}{} & 
\end{tabular}
$$
where the sets $I^a, V_2^a, I^\sim, V_2^\sim$ may be empty and $D(a, P)=I^a\cup V_2^a\cup V_1$. Similarly we can do for a profile $P\in \cP_b(\ell, k)$. Therefore, the following proposition is proved.

\begin{prop}\label{13}
The following equations hold true:
$$\cP_a (\ell, k) = \bigcup_\sigma\,\, \cP_a(\sigma(I), \sigma(\cF)),$$
$$\cP_b (\ell, k) = \bigcup_\sigma\,\, \cP_b(\sigma(I), \sigma(\cF)),$$
where: $\sigma$ runs over all permutations of $V,$
$I$ is a subset of $V$ with cardinality $\ell$, $\cF$ is the SSCF  on $I^c$
of cardinal type $k$ (with respect to $I^c$).
\end{prop}

Since $\cG_{\ell+k}\subseteq \cG_k$ and dually $\cG_{\ell+k}^\circ\supseteq \cG_k^\circ$, we also have the next proposition.

\begin{prop}\label{14}
$$P\in \cP_a (\ell, k) \Leftrightarrow \left\{ 
\begin{array} {ll}
D(a,P) \in \cG_k ,   & \\

D(b,P)\notin\cG_{\ell+k}^\circ   & \\
\end{array}
\right.\Leftrightarrow P\in \cP_a(\cG_k)\setminus\cP_b(\cG_{\ell+k}),$$

$$P\in \cP_b (\ell, k) \Leftrightarrow \left\{ 
\begin{array} {ll}
D(a,P) \notin \cG_k ,   & \\

D(b,P)\in\cG_{\ell+k}^\circ   & \\
\end{array}
\right.\Leftrightarrow P\in \cP_b(\cG_{\ell+k})\setminus\cP_a(\cG_{k}).$$
\end{prop}

We conclude by presenting the definition of a scf that will be useful for an intermediated step in the completion of the proof of Theorem \ref{first}. Given a sequence ${ \langle\cccell, \ccck}\rangle=(\ell_\lambda, k_\lambda)$, for $\lambda \in \Lambda =\{ 0, 1, 2 \dots, |\Lambda|\}$,  we say that a profile $P$ has index $\lambda(P)$ if $\lambda(P)$ is the smallest index $\lambda$ for which $P\in\cP_\lambda\overset{def}=\cP(\ell_\lambda, k_\lambda)$. This notion replicates, mutatis mutandis, the one preceding Definition \ref{extended} (see also \cite[Definitions 3.1 and 3.2]{KPS2}).

\begin{defin}\label{def14}
Given $x\in\{a, b\}$, and the sequence ${ \langle\cccell, \ccck}\rangle=(\ell_\lambda, k_\lambda)$, where $\lambda \in \Lambda =\{ 0, 1, 2 \dots, |\Lambda|\}$,  
we define the following scf: 

 \begin{equation*}
\psi_{{ \langle\ccell, \cck}\rangle,\, x}(P) = \left\{ 
\begin{array} {ll}
 a,   & \mbox{ if }  P\in \cP_a(\ell_{\lambda(P)}, k_{\lambda(P)}) \\

 b,   & \mbox{ if }  P\in \cP_b(\ell_{\lambda(P)}, k_{\lambda(P)}) \\
 x,   & \mbox{ if } P\notin  (\bigcup_{\lambda\in \Lambda} \cP_\lambda).\\
\end{array}
\right.
\end{equation*}
\end{defin}

\begin{oss}\label{36}
{\sl With reference to  Definition \ref{def14}, if we adopt, for $\lambda=0,1,2, ..., |\Lambda|$, the following shorter notation:
\quad $\psi_\lambda\overset{def}=\psi_{\langle (\ell_0, \dots, \ell_\lambda), (k_0, \dots, k_\lambda)     \rangle, x}$, \quad we can notice that, obviously,

on $\cP_0$ we have $\psi_0=\psi_\lambda, \,\,\forall \lambda\ge 0$. \quad
On $\cP_1\setminus\cP_0$ we have $\psi_1=\psi_\lambda, \,\,\forall \lambda\ge 1$.\,\,
...
\,\,
On $\cP_\alpha\setminus (\bigcup _{\beta<\alpha}\cP_\beta)$ we have $\psi_\alpha=\psi_\lambda, \,\,\forall \lambda\ge \alpha$;
\quad
On $\cP\setminus (\bigcup _{\lambda\in\Lambda}\cP_\lambda)$ we have $\psi_\lambda(\cdot)=x, \,\,\forall \lambda\in \Lambda.$
}\end{oss}

For the definition of the SCF $\phi_{\ccG,\, x}$ employed in the following proposition we refer to \cite[Remark 3.3]{KPS2}.

\begin{prop}\label{teo18}
Assume for the  sequences $(\ell_\lambda, k_\lambda)_{\lambda\in \Lambda}$ $(with \,\,\Lambda =  \{0,1, \dots, |\Lambda|\})$ that $\ccck$ is decreasing and $\cccell+\ccck$ is increasing.%\footnote{Namely the sequence $\cccm$ is decreasing.}

The scf $\psi_{{ \langle\ccell, \cck}\rangle,\, x}$
 is identical to the scf $\phi_{\ccG,\, x}$ where the collection $\cccG$ is:
 
 \begin{equation*}
 \cccG=\left\{ 
\begin{array} {ll}
\cG_{k_0}, \cG_{\ell_0+k_0}, \cG_{k_1}, \cG_{\ell_1+k_1}, \cG_{k_2}, \cG_{\ell_2+k_2},\dots,  \cG_{k_{|\Lambda|}}, \cG_{\ell_{|\Lambda|}+k_{|\Lambda|}},   & if x=a \\
 \\
\cG_{\ell_0+k_0}, \cG_{k_0}, \cG_{\ell_1+k_1}, \cG_{k_1}, \cG_{\ell_2+k_2}, \cG_{k_2}, \dots, \cG_{\ell_{|\Lambda|}+k_{|\Lambda|}}, \cG_{k_{|\Lambda|}},   & if x=b\\
\end{array}
\right.
\end{equation*}

being the committees constituting $\cccG$,  to be considered in the order from left to right described in the above formula. 
\end{prop}

\begin{oss}\label{28}{\sl Before giving the proof we observe that we are saying, on the base of  Remark \ref{nest}, that $\psi_{{ \langle\ccell, \cck}\rangle,\, x}$ is the extended quota majority method defined: 

by the sequence 
\quad 
$({k_0}, {\ell_0+k_0}, {k_1}, {\ell_1+k_1}, {k_2}, {\ell_2+k_2},\dots,  {k_{|\Lambda|}}, {\ell_{|\Lambda|}+k_{|\Lambda|}}, 0)$, if $x=a$,

by the sequence 
\quad
$({\ell_0+k_0}, {k_0}, {\ell_1+k_1}, {k_1}, {\ell_2+k_2}, {k_2}, \dots, {\ell_{|\Lambda|}+k_{|\Lambda|}}, {k_{|\Lambda|}}, n+1)$  if $x=b$.}\eproof\end{oss}

\dimo
Let us shorten the notation for the two functions, by using simply $\psi$ and $\phi$. Also,  according to our notation, $m$ (possibly indexed, as it is now) is $n-\ell-k+1$.

We show that for every profile $P$, we have
$\phi(P)=\psi(P).$

Notice that 

$$\cP_a(\cG_{\ell+k})\subseteq \cP_a(\cG_k); \mbox{\quad  and \quad}\cP_b(\cG_{k})\subseteq \cP_b(\cG_{\ell+k}).$$
%Moreover, remind that from Proposition \ref{14} we have 
%$$\cP_a(\ell,k)=\cP_a(\cG_k)\setminus \cP_b(\cG_{\ell+k})\mbox{\quad  and \quad}\cP_b(\ell,k)=\cP_b(\cG_{\ell+k})\setminus \cP_a(\cG_{k})
%$$

To prove that the two scfs coincide on $P$ we distinguish two cases. Let us consider first the case that 
$P\in\bigcup_{\lambda\in\Lambda}\cP_\lambda$. Suppose that $\lambda$ is the index of the profile $P$. We have either $P\in \cP_a(\ell_\lambda, k_\lambda)$  or  $P\in \cP_b(\ell_\lambda, k_\lambda)$, respectively giving $\psi(P)=a$ or $\psi(P)=b$. We show that correspondingly $\phi$ attains the same value on $P$.  

\smallskip
Case $P\in \cP_a(\ell_\lambda, k_\lambda)$.

Since by Proposition \ref{14}, $ \cP_a(\ell_\lambda, k_\lambda)= \cP_a(\cG_{k_\lambda})\setminus\cP_b(\cG_{\ell_\lambda+k_\lambda})\subseteq 
[\cP_a(\cG_{\ell_\lambda+k_\lambda})\cup \cP_a(\cG_{k_\lambda})]\setminus\cP_b(\cG_{\ell_\lambda+k_\lambda})$,   if we prove that $P$ does not belong to $\cP(\cG_{\ell_\beta+k_\beta})\cup\cP(\cG_{k_\beta})$ whenever $\beta<\lambda$, we shall have that $\phi(P)=a$. In other words we have to exclude that $P$ belongs to one of the following four sets:
$\cP_a(\cG_{\ell_\beta+k_\beta})\subseteq\cP_a(\cG_{k_\beta})$,
\quad
$\cP_b(\cG_{k_\beta})\subseteq \cP_b(\cG_{\ell_\beta+k_\beta})$.

We show that $P\notin \cP_b(\cG_{\ell_\beta+k_\beta})$.

Suppose the contrary, then $|D(b,P)|\ge m_\beta$. On the other hand since $P\notin\cP_b(\cG_{\ell_\lambda +k_\lambda})$, we have $|D(b,P)|< m_\lambda$ against the assumption that $\cccm$ is decreasing.

Now we can show that $P\notin \cP_a(\cG_{k_\beta})$. Indeed, if not by Proposition \ref{14} we have $P\in \cP_a(\ell_\beta+k_\beta)$, against the definition of index.

\bigskip
Case $P\in \cP_b(\ell_\lambda, k_\lambda)$.

By Proposition \ref{14} $P\in [\cP_b(\cG_{\ell_\lambda+k_\lambda})\cup \cP(\cG_{k_\lambda})]\setminus \cP_a(\cG_{k_\lambda})$. An argument similar to the previous one applies. If the index $\beta$ is less than $\lambda$, the profile $P$ cannot belong to $\cP_a(\cG_{k_\beta})$ otherwise the  monotonicity of $\ccck$ is violated. Now, the profile cannot be in $\cP_b(\cG_{\ell_\beta+k_\beta})$, otherwise it belongs to $\cP_b(\ell_\beta+k_\beta)$, against the definition of index.

\bigskip
It remains to consider the case that $P\notin\bigcup_{\lambda\in\Lambda}\cP_\lambda$.

In this case the value of $\psi(P)$ is $x$, and we show that even $\phi(P)$ is $x$. We have to investigate the case that $P\in \bigcup_{\lambda\in\Lambda} [\cP(\cG_{k_\lambda})\cup\cP(\cG_{\ell_\lambda+k_\lambda})]$, since in the opposite case the assertion comes from the definition of $\phi$. So, let us suppose that  $P\in \bigcup_{\lambda\in\Lambda} [\cP(\cG_{k_\lambda})\cup\cP(\cG_{\ell_\lambda+k_\lambda})]$ and let $\alpha$ be the first (moving from left to right in the definition of $\cccG$) index such that $P\in \cP(\cG_\alpha)$. We are done if we show that 
$$x=a\Rightarrow P\notin \cP_b(\cG_\alpha).$$
Indeed, if this is not the case, 
$P\in \cP_b(\cG_\alpha).$ But $\alpha$ is either some $k_\lambda$ or an $\ell_\lambda+k_\lambda$ and in both cases we can write 
$P\in \cP_b(\cG_{\ell_\lambda+k_\lambda}).$ Since $P\notin\bigcup_{\lambda\in\Lambda}\cP_\lambda$, by Proposition \ref{14} we must have $P\in \cP_a(\cG_{k_\lambda}).$ Hence $\alpha=k_\lambda$ and $P\in \cP_a(\cG_\alpha)\cap \cP_b(\cG_\alpha),$ a contradiction.

Similarly, we can  show that 
$$x=b\Rightarrow P\in \cP_b(\cG_\alpha),$$
and we are done.
\eproof

\subsection{Anonymous, non manipulable  scfs are extended quota majority methods}\label{prova1}
\lhead{\sc \scriptsize Representation of  anonymous...}
\rhead{\sc  \scriptsize Anonymous, non manipulable  is extended quota majority}
First we see that committees of cardinal type arise when anonymous scfs are considered.
Compare indeed next proposition with \cite[Proposition 4.1]{KPS2}.  We recall that the committees of cardinal type on $V$ are $\cccV=\{\cG_k: k=1,\dots, n\}$, $\cG_k=\{E\subseteq V: |E| \geq k\}$. We also consider  the power set of $V$ and the empty subset of the power set of $V$  as committees of  cardinal type   respectively zero ($\cG_0=2^V$) and $n+1$ ($\cG_{n+1}=\O$).

\begin{prop}\label{truetrue}
Let $\phi$ be a non-manipulable  scf. If $\phi$ is anonymous,  the (unique)
committee $\cal F$ such that for every profile $P$ one has

$$ D(a, P)\in {\cal F}\Rightarrow \phi(P)= a, \quad \mbox{ and } \quad
 D(b, P)\in {\cal F}^\circ\Rightarrow \phi(P)= b,$$
is of cardinal type.

\end{prop}
\dimo

If we denote by $\vr(\cF)$ the least cardinality of the coalitions belonging to $\cF$, we have to prove that 

$$|F|\geq \vr(\cF) \Rightarrow F\in \cF.$$
Let $D\in\cF$ be such that $|D|=\vr(\cF).$

Let $E\subseteq F$ be such that $|D|=|E|$. Then take a permutation $\sigma$ of $V$  such that the $\sigma$-image of $E$ is $D$.

Take a strict profile $S$ that ranks $a$ as top on $D$  and $b$ as top on $E^c$. We hence have $\phi(S)=a$. 

Let $Q$ be the profile $S\circ \sigma$, i.e. $Q_v=S_{\sigma(v)}.$ Because of anonymity and since $Q$ is strict, we have that
$D(a, Q)\in \cF$.

Clearly $v\in D(a,Q)$ is same as $\sigma(v)\in D(a,S)=D$, which is the same as $v\in E$. So $E$ is in $\cF$ and therefore the superset $F$ too is in $\cF$.
\eproof

In the above proposition  $\cF\in\cccV$ when $\phi$ is onto. $\cF\in \{\cG_0, \cG_{n+1}\}$ for the constant scfs $a$ and $b$ respectively.

The following is the crucial step to obtain the results presented in Section \ref{risultati}.

\begin{theo}\label{teo17}
Let $\phi$ be a scf which is onto, anonymous, and non-manipulable. Say $x$ is the collective choice corresponding to the unanimous indifference.Correspondingly, we can find a sequence ${ \langle\cccell, \ccck}\rangle=(\ell_\lambda, k_\lambda)_{\lambda\in\Lambda}$ such that:
\begin{enumerate}
\item  the sequence $\cccell$ is strictly increasing and $\ell_0=0,$\;
\item in case $x=b$ (respectively, $x=a$), the sequence $\ccck$ is strictly decreasing (respectively, decreasing) and the sequence $\cccell+\ccck$ is increasing (respectively, strictly increasing).
\item $\phi=\psi_{{ \langle\ccell, \cck}\rangle,\, x}$.
\end{enumerate}
\end{theo}

\dimo
For the  proof, we shall go along steps that  echo those  for proving \cite[Theorem 4.2]{KPS2}.
However, a deeper argument  is needed to achieve the result.

We remind that once the sequences $\cccell$ and $\ccck$ are given, the sequence $\cccm$ is also fixed for the dual committees. We also remind the notation introduced in Remark \ref{36}.

Due to Proposition \ref{truetrue}, let $\cF_0$ be the SSCF on $V$ of cardinal type, say $k_0$, such that 
$$ |D(a, P)|\ge k_0 \Rightarrow \phi(P)= a, \quad\mbox{ and } \quad |D(b, P)|\ge n-k_0+1\Rightarrow \phi(P)= b.$$
Set: \quad  
$I_0=\O$, $\ell_0=|I_0|$, $m_0=n-\ell_0-k_0+1$, 
$\cP_0=\cP(\ell_0,k_0)$. 

Clearly on $\cP_0$ we have that $\phi=\psi_{\langle \ell_0, k_0\rangle, x}=\psi_0$, hence, if the set of profiles $\{P\notin \cP_0: \phi(P)\neq x\}$ is empty, the theorem is proved, being $\langle \ell_0, k_0\rangle$ the desired sequence. So, let us suppose it is nonempty and let $P^1$ be one of its members with $I(P^1)=:I_1$ of smallest cardinality, that we  denote by $\ell_1$. By definition $1\le\ell_1<n$.

\smallskip
Applying Proposition \ref{truetrue} to the scf defined for the society $I_1^c$ as follows:

 $$P_{I_1^c}\longrightarrow \phi([\sim_{I_1}, P_{I_1^c}]),$$ 
we determine a SSCF $\cF_1$ on $I_1^c$ of cardinal type (with respect to $I_1^c$), say $k_1$  , such that, with $m_1=n-\ell_1-k_1+1$,

$$(+)\quad  |D(a, P)\cap I_1^c|\ge k_1 \Rightarrow \phi([\sim_{I_1},P_{I_1^c}])= a, \mbox{ and }  |D(b, P)\cap I_1^c|\ge m_1\Rightarrow \phi([\sim_{I_1},P_{I_1^c}])= b.$$
Set $\cP_1=\cP(\ell_1, k_1)$. 

Notice that if for a profile $P$ we have $I(P)=I_1$, necessarily it is true that $P\in \cP_1$. 

Since $P^1\in \cP_1\setminus \cP_0$, we must have $|D(a, P^1)|<k_0$ and $|D(b, P^1)|<m_0$.

\bigskip

To fix ideas, let us assume, throughout the sequel, that $x=b$. The argument for the case $x=a$ is the same, except for reversing the role of the the sequences $\ccck$ and $\cccm$. 

\bigskip
Since $\phi(P^1)=a$ we have that necessarily $|D(a, P^1)|\ge k_1$ and therefore $$k_1<k_0.$$

Let $Q^1$ be a profile identical to $P^1$ on $I_1$, identical to $P^1$ on a subset of $D(a,P^1)$ with cardinality $k_1$, and reporting $b$ as the top choice of  the remaining $m_1-1$ voters.

For the new profile $Q^1$ we have $\phi(Q^1)=a$, and also $Q^1\in\cP_1\setminus \cP_0$. The latter implies\footnote{If we assume that $Q^1\in \cP_0$, then $\psi_0(Q^1)=a$. Hence necessarily $Q^1\in\cP_a(\ell_0, k_0)$. This leads to $k_1\ge k_0$ wich is false. 

\noindent
Having seen that  $Q^1\notin \cP_0$, also $Q^1\notin \cP_b(\ell_0, k_0)$, namely we have $|D(b,Q^1)|<m_0$ hence $m_1-1\le m_0-1$.  } that 
$$m_1\le m_0.$$
CLAIM:  on $\cP_0\cup\cP_1$ the scfs $\phi$ and the scf  $\psi_{\langle\ccell, \cck\rangle , x}$ ($=\psi_1$), where $\cccell= (\ell_0, \ell_1)$ and $\ccck=(k_0, k_1)$, coincide.

Indeed, let us consider a profile $P\in\cP_1\setminus \cP_0$.  We have to show that

$$P\in \cP_a(\ell_1, k_1) \Rightarrow \phi(P)=a \mbox{ and } P\in \cP_b(\ell_1, k_1) \Rightarrow \phi(P)=b.$$ 
We show the first implication only.
Applying Proposition \ref{13} we can find a permutation $\sigma$ of $V$ such that $P\in \cP_a(\sigma(I_1), \sigma(\cF_1))$, namely such that the profile  $(P\circ\sigma)$  belongs to $ \cP_a(I_1, \cF_1)$.
By Definition \ref{13d} 2., $(+)$ and strategy-proofness,
%\footnote{See the proof of \cite[Theorem 4.2]{KPS2}}, 
 we have that $\phi(P\circ\sigma)=a$. By anonymity we get $\phi(P)=a$.

Now it is clear that if the set of profiles $\{P\notin (\cP_0\cup\cP_1): \phi(P)\neq x\}$ is empty, the theorem is proved, being  $\cccell= (\ell_0, \ell_1)$ and $\ccck=(k_0, k_1)$ the desired sequences.

\bigskip
If the set $\{P\notin (\cP_0\cup\cP_1): \phi(P)\neq x\}$ is nonempty, we shall  apply the Lemma \ref{18} below, which is really the induction step.We apply repeatedly Lemma \ref{18} till we stop, that is, when $\{P\notin (\cP_0\cup\dots\cup\cP_r): \phi(P)\neq x \}$ is empty reaching the desired representation. \eproof

%So we have to investigate the alternative case. For doing this, 
%Presently we apply the mentioned Lemma with reference to $r=2$, and, in case the set of profiles $\{P\notin (\cP_0\cup\cP_1\cup\cP_2): \phi(P)\neq x \}$ is nonempty, we apply it again for $r=3$ and so on. 

\begin{lem}\label{18}
Let $\phi$ be a scf which is onto, anonymous, non-manipulable, and assigns $x\in \{a, b\}$ to the profile in which all agents are indifferent. Let the sequences  $\ell_\lambda, k_\lambda$ and $m_\lambda$ (for $\lambda\in\{0, 1, \dots, r-1\}$)  be  such that $$0\leq\ell_\lambda<n, \qquad 1\leq k_\lambda\leq n-\ell_\lambda,\qquad m_\lambda=n-\ell_\lambda-k_\lambda+1.$$

 Assume further that $\ell_0=0,$ $\ell_\lambda$ is strictly increasing, and 
\begin{itemize}
\item [A1] for every $1\le \lambda\le r-1$,  there is a profile $P^\lambda\in \cP_\lambda \setminus  (\cP_0\cup\dots \cup\cP_{\lambda-1})$ with:  $\phi(P^\lambda)\neq x$; $I_\lambda:=I(P^\lambda)$; $\ell_\lambda=|I_\lambda|$  
$ = \min \{|I(P)|: P\notin (\cP_0\cup\dots \cup\cP_{\lambda-1}), \phi(P)\neq x\}$, 

\item [A2]  in case $x=b$ (respectively, $x=a$), $k_\lambda$ is strictly decreasing, $m_\lambda$ is decreasing
(respectively, $k_\lambda$ is decreasing, $m_\lambda$ is strictly decreasing)

\item [A3] $\phi=\psi_{r-1}$ on $\cP_0\cup \dots\cup \cP_{r-1}$

\item [A4] $\{P\notin (\cP_0\cup\dots \cup\cP_{r-1}): \phi(P)\neq x\}$ is nonempty.
\end{itemize}

Then we can find  $\ell_r>\ell_{r-1}$,\, $k_r<k_{r-1}$,\, $m_r\le m_{r-1}$ (in case $x=b$, otherwise for $x=a$ we shall have $k_r\le k_{r-1}$,\, $m_r< m_{r-1}$), and a profile $P^r\in \cP_r \setminus  (\cP_0\cup\dots \cup\cP_{r-1})$ with $\phi(P^r)\neq x$, such that 
$$|I(P^r)|=\ell_r = \min \{|I(P)|: P\notin (\cP_0\cup\dots \cup\cP_{r-1}), \phi(P)\neq x\},
\mbox{ and }$$ $$\phi=\psi_{r} \mbox{ on } \cP_0\cup \dots\cup \cP_{r}.$$

\end{lem}

\dimo
Notice that whenever $P$ is a profile with $I(P)=I_\lambda$,  we necessarily have $P\in \cP_\lambda$. 

Now, using {\it A4},  let $P^r$ be a profile in $\{P\notin (\cP_0\cup\dots \cup\cP_{r-1}): \phi(P)\neq x\}$  with $I(P^r)=:I_r$ of smallest cardinality, that we  denote by $\ell_r$. By definition $\ell_{r-1}\le\ell_r<n$.

\bigskip
CLAIM 1: \qquad $\ell_{r-1}<\ell_r.$

Suppose the contrary, i.e. that $\ell_{r-1}=\ell_r$.  Let $\sigma$ be a permutation of $V$ that maps $I_{r-1}$ onto $I_r$. Since for  the profile $P^r\circ\sigma$ we have  $I(P^r\circ\sigma)=I_{r-1}$, certainly we also have that $P^r\circ\sigma\in \cP(\ell_{r-1}, k_{r-1})$. Hence $P^r\in \cP(\ell_{r-1}, k_{r-1})$, against the definition of $P^r$.\eproof

\bigskip
Applying Proposition \ref{truetrue} to the scf defined for the society $I_r^c$ as follows:

 $$P_{I_r^c}\longrightarrow \phi([\sim_{I_r}, P_{I_r^c}]),$$ 
we determine a SSCF $\cF_r$ on $I_r^c$ of cardinal type  $k_r$ (with respect to $I_r^c$) such that, with $m_r=n-\ell_r-k_r+1$,

$$(+)\quad  |D(a, P)\cap I_r^c|\ge k_r \Rightarrow \phi([\sim_{I_r},P_{I_r^c}])= a, \mbox{ and }  |D(b, P)\cap I_r^c|\ge m_r\Rightarrow \phi([\sim_{I_r},P_{I_r^c}])= b.$$
Set $\cP_r=\cP(\ell_r, k_r)$. 

Notice that whenever we have $I(P)=I_r$, necessarily it is true that $P\in \cP_r$. Hence $P^r\in \cP_r\setminus (\cP_0\cup\dots\cup \cP_{r-1})$.

\bigskip
Claims 2 and 3 below, refers to $x=b$. For the case $x=a$ just reverse the role of $k$'s and $m$'s first proving 
$m_r<m_{r-1}$, then $k_r\le k_{r-1}$.

\bigskip
CLAIM 2: \qquad $k_r< k_{r-1}$.

Let us suppose on the contrary that $k_r\ge k_{r-1}$. Since $\phi(P^r)=a$, we have necessarily that $|D(a, P^r)|\ge k_r$. Hence we have $|D(a, P^r)|\ge k_{r-1}$. Showing also that $|D(b, P^r)|< m_{r-1}$, we have $P^r\in\cP_{r-1}$ which is a contradiction. For our purpose it is enough to use  CLAIM 1. Indeed we can write:  $|D(b, P^r)|\le n-\ell_r- k_r<n-\ell_{r-1}- k_{r-1}< m_{r-1}$.\eproof

\bigskip
CLAIM 3: \qquad $m_r\le m_{r-1}$.

Again by contradiction assume that $m_r > m_{r-1}$.
 
Let $Q^r$ be a profile identical to $P^r$ on $I_r$, identical to $P^r$ on a subset of $D(a,P^r)$ with cardinality exactly $k_r$, and reporting $b$ as the top choice of  the remaining $m_r-1$ voters.

For the new profile $Q^r$ we have $\phi(Q^r)=a$. 

Now, since $m_r-1 \ge m_{r-1}$, we have $|D(b, Q^r)|=m_r-1\ge m_{r-1}$. But also, $|D(a, Q^r)|=k_r<k_{r-1}$, hence $Q^r\in\cP_b(\ell_{r-1}, k_{r-1})$. Because of the monotonicity of $\ccck$, for every $\lambda<r-1$, the profile  $Q^r\notin \cP_a(\ell_\lambda, k_\lambda)$.  Hence, by the assumption {\it A3} we have that $\phi(Q^r)=b$, a contradiction.\eproof

\smallskip
We shall finally prove that on the set  $  \cP_0\cup \dots\cup \cP_{r}$ of profiles, the two scfs 
$\phi$ and $\psi_{r}$ are identical. 

For this purpose, let us consider a profile $P\in\cP_r\setminus (\cP_0\cup\dots\cup \cP_{r-1})$.  We have to show that 
$$P\in \cP_a(\ell_r, k_r) \Rightarrow \phi(P)=a \mbox{ and } P\in \cP_b(\ell_r, k_r) \Rightarrow \phi(P)=b.$$ 
We show the first implication only.
Applying Proposition \ref{13} we can find a permutation $\sigma$ of $V$ such that $P\in \cP_a(\sigma(I_r), \sigma(\cF_r))$, namely such that the profile  $(P\circ\sigma)$  belongs to $ \cP_a(I_r, \cF_r)$.
By Definition \ref{13d}, $(+)$ and strategy-proofness, we have that $\phi(P\circ\sigma)=a$. By anonymity we get $\phi(P)=a$.
\eproof

%\bigskip
%punto

{\bf Completion of the proof of Theorems \ref{first} and \ref{main}}:

Now, if we start with an onto, anonymous, non-manipulable, binary scf $\phi$, applying Theorem \ref{teo17}, Proposition \ref{teo18}, and Remark \ref{28} in this sequence  gives us the proof of Theorem \ref{first}. 

Theorem \ref{main} follows from Proposition \ref{33}.\eproof

\begin{oss}\label{sempre per bhaskara}
{\sl
If in Theorem \ref{teo17} we had that the sequence $\cccell+\ccck$ is  strictly increasing, we would have obtained the proper representation directly by means of Proposition \ref{teo18}. But in general, it is possible that $\cccell+\ccck$ is not strictly increasing. A relevant example is the scf $\phi$ of Example \ref{per bhaskara}.

According to Proposition \ref{teo18}, from the procedure of the proof of Theorem \ref{teo17}, we get  for the mentioned $\phi$ the sequence $(5,5, 5, 4, 5,3,5,2,12)$, where $\cccell+\ccck$ is constantly 5. We shall show this below.

\bigskip
{\sc Computing $\cccell$ and $\ccck$}:
$\ell_0=0$ and $k_0=5$ are easily obtained by the restriction to strict profiles. It is obvious that $\{P\notin\cP_0: \phi(P)\neq b\}$ is nonempty and consists of the profiles $P$ for which $2\le |D(a,P)|< 5  \,\&\, |D(b,P)|<7$. Therefore  $\ell_1=1$.

Considering the restriction $\phi(\sim_1, P_{\{2, \dots, 11\}})$ on the society $\{2, \dots, 11\}$, we get that $k_1=4$, and consequently $m_1=7$.

Hence, $\{P\notin \cP_0\cup\cP_1: \phi(P)\neq b\}$ consists of the profiles $P$ for which $2\le |D(a,P)|< 4  \,\&\, |D(b,P)|<7$. Therefore  $\ell_2=2$.

Considering the restriction $\phi(\sim_{\{1,2\}} P_{\{3, \dots, 11\}})$ on the society $\{3, \dots, 11\}$, we get that $k_2=3$, and $m_2=7$. 

Hence, $\{P\notin \cP_0\cup\cP_1\cup \cP_2: \phi(P)\neq b\}$ consists of the profiles $P$ for which $2\le |D(a,P)|< 3  \,\&\, |D(b,P)|<7$.  We conclude by determining $\ell_3=3, k_3=2, m_3=7.$\eproof

The proper representation of $\phi$ is given by (5,2,12).}
\end{oss}

%\newpage

{\color{black}\section{Comparison with Lahiri and Pramanik representation}
\lhead{\sc \scriptsize Representation of  anonymous...}
\rhead{\sc  \scriptsize Comparison  }

In \cite{LP}, Lahiri and Pramanik show that the anonymous, onto, non-manipulable scfs are (all and only)  {\it quota rules} either {\it with indifference default $a$}, denoted by $f^{r, \ccx}_a$, or {\it with indifference default $b$}, denoted by $f^{r, \ccy}_b$.

In the above notation Lahiri and Pramanik assume that $r\in\{1, 2, \dots, n\},\,\, \cccx, \cccy$ are $r$-dimensional vectors such that 

\begin{itemize}
\item $x_i\le x_{i+1}\le x_i+1$, for all $i\in\{1,\dots, r-1\}$
\item $\cccx\in \{1\}\times\{1,2\}\times\dots\times\{1, \dots r\}$
\item $y_i\le y_{i+1}\le y_i+1$, for all $i\in\{1,\dots, r-1\}$
\item $\cccy\in \{(n-r)+1\}\times\{(n-r)+1, (n-r)+2\}\times\dots\times\{(n-r)+1, (n-r)+2, \dots n\}$.
\end{itemize} 

The quota rules can be described as in the table below  (compare with \cite[Definitions 10 and 11]{LP}) 

$$\begin{tabular}{|c|c|c|}
\hline
{\footnotesize Cardinality of} $I(P)$ & {\footnotesize Collective choice} $f_{a}^{r,\ccx}(P)$ & {\footnotesize Collective
choice} $f_{b}^{r,\ccy}(P)$ \\ \hline
$r$, {\footnotesize or more} & $a$ & $b$ \\ \hline
$r-1$ & $a,$ if $|D(a,P)|\geq x_{1}$, {\footnotesize otherwise} $b$ & $b,$ if $|D(b,P)|\geq
y_{1}^{\prime }$, {\footnotesize otherwise} $a$ \\ \hline
$r-2$ & $a,$ if $|D(a,P)|\geq x_{2}$, {\footnotesize otherwise} $b$ & $b,$ if $|D(a,P)|\geq
y_{2}^{\prime }$, {\footnotesize otherwise} $a$ \\ \hline
$r-3$ & $a,$ if $|D(a,P)|\geq x_{3}$, {\footnotesize otherwise} $b$ & $b,$ if $|D(a,P)|\geq
y_{3}^{\prime }$, {\footnotesize otherwise} $a$ \\ \hline
... & ... & ... \\ \hline
$r-i$ & $a,$ if $|D(a,P)|\geq x_{i}$, {\footnotesize otherwise} $b$ & $b,$ if $|D(a,P)|\geq
y_{i}^{\prime }$, {\footnotesize otherwise} $a$ \\ \hline
... & ... & ... \\ \hline
$1$ & $a,$ if $|D(a,P)|\geq x_{r-1}$, {\footnotesize otherwise} $b$ & $b,$ if $|D(a,P)|\geq
y_{r-1}^{\prime }$, {\footnotesize otherwise} $a$ \\ \hline
$0$ & $a,$ if $|D(a,P)|\geq x_{r}$, {\footnotesize otherwise} $b$ & $b,$ if $|D(a,P)|\geq
y_{r}^{\prime }$, {\footnotesize otherwise} $a$ \\ \hline
\end{tabular}
$$
where, for $1\le i\le r$, we have set, for symmetry,  $y'_i =(n-r+1)-y_i+i\in\{i, i-1, \dots, 1\}.$ Notice that, similarly to $\cccx$, for $\cccy'$ we have
\begin{itemize}
\item $\cccy'\in \{1\}\times\{1,2\}\times\dots\times\{1, \dots r\}$ (and
$y'_i\le y'_{i+1}\le y'_i+1$, for all $i\in\{1,\dots, r-1\}$).
\end{itemize} 
Also observe that $x_r$ corresponds to $k_0$ in our representation.

The difference between Lahiri and Pramanik representation and ours  is evident. Moreover the final example shows that our representation theorem is simpler, involving a smaller number of parameters.

Given a scf $\phi$, anonymous and non-manipulable, after determining the indifference quota $r$ that indisputably gives rise to the default collective choice, Lahiri and Pramanik have to  discuss all cases $r>|(I(P)|\ge 0$, by means of the further parameters
$x_{[r-|I(P)|]}$ in case default is $a$ ( $y'_{[r-|I(P)|]}$, for default $b$).

Using an approach different from that of Lahiri and Pramanik \cite{LP}, we have provided an algorithm that produces a unique, up and down, sequence of majority quotas, that applied in the given order gives back the scf $\phi$.

{\bf Example \ref{per bhaskara} continued.}
For the scf  $\phi$ of this example, we know that our  proper representation formula gives $\phi=\phi_{(5,2,12)}$. Let us determine the representation of $\phi$ as $\phi=f_{b}^{r,\ccy}$, according to Lahiri and Pramanik \cite{LP}. We need to determine $r$ and the vector $\cccy$.

It is clear that $r=10$. Hence we have that $\cccy$ has dimension 10, and, for $i\in \{1,\dots, 10\}$, we have $y_i=i+2-y'_i$ where
$$(y'_1, y'_2, \dots, y'_7)=(1,2,3,4,5,6,7) \mbox{ and } (y'_8, y'_9, y'_{10})=(7,7,7).
$$
Finally $$\cccy=(2,2,2,2,2,2,2,3,4,5).$$
Notice that other choices for $(y'_8, y'_9, y'_{10})$ are possible, so that the Lahiri and Pramanik representation is not unique, contrary to ours.
}

\end{document}